\documentclass[preprint,prd,twocolumn,tightenlines,floatfix,showpacs,byrevtex,nofootinbib]{revtex4}
\usepackage{pictex}
\usepackage[dvips]{graphicx}
\usepackage{amsmath}

\newcommand{\eqn}[1]{ \begin{equation} #1 \end{equation} }
\newcommand{\four}{\!\!\!\!/}

\newcommand{\half}{\frac{1}{2}}

\begin{document}
\preprint{ADP-02-83/T522}

\title{Generalised Spin Projection for Fermion Actions}

\author{Waseem Kamleh}
\email{wkamleh@physics.adelaide.edu.au}
\affiliation{CSSM Lattice Collaboration, Special Research Centre for the Subatomic Structure of Matter and Department of Physics and Mathematical Physics, University of Adelaide 5005, Australia. }

\begin{abstract}
The majority of compute time doing lattice QCD is spent inverting the fermion matrix. The time that this takes increases with the condition number of the matrix. The FLIC(Fat Link Irrelevant Clover) action displays, among other properties, an improved condition number compared to standard actions and hence is of interest due to potential compute time savings. However, due to its two different link sets there is a factor of two cost in floating point multiplications compared to the Wilson action. An additional factor of two has been attributed due to the loss of the so-called spin projection trick.  We show that any split-link action may be written in terms of spin projectors, reducing the additional cost to at most a factor of two. Also, we review an efficient means of evaluating the clover term, which is additional expense not present in the Wilson action.
\end{abstract}

\maketitle

\section{Introduction}

The FLIC(Fat Link Irrelevant Clover) action\cite{zanotti-hadron} has become of interest recently as an alternative to standard actions (such as Wilson or Clover) due to its superior condition number\cite{kamleh-overlap}. This allows for more efficient fermion matrix inversion\cite{zanotti-hadron}, which is used in the calculation of propagators and dynamical configurations, and in evaluating the matrix sign function in the overlap fermion formalism\cite{neuberger-practical, edwards-practical}. Hence, actions with an improved condition number have the potential to save significant compute time.

To begin, we review the spin-projection trick\cite{alford-improving} for the Wilson action, which utilises projection operators in spinor space to reduce the computation required to evaluate the Wilson action. We then generalise this trick to the broader class of split-link actions. Finally, we examine the FLIC action specifically, and discuss a similar trick for reducing the cost of evaluating the clover term.

\section{Standard Spin-Projection Trick}

The Wilson operator
\eqn{D_{\rm w} = \nabla\four + \frac{1}{2}\Delta + m}
can be written as
\begin{multline}
(D_{\rm w}\psi)_x =  (4+m)\psi_x-\half\sum_\mu (1-\gamma_\mu) U_\mu(x)\psi_{x+\mu} \\ + (1+\gamma_\mu) U^\dagger_\mu(x)\big)\psi_{x-\mu} \\
  =  (4+m)\psi_x-\sum_\mu U_\mu(x)\Gamma_\mu^- \psi_{x+\mu} \\ + U^\dagger_\mu(x)\Gamma_\mu^+ \psi_{x-\mu},
\end{multline}
where we have defined the spin projectors
\eqn{\Gamma_\mu^\pm = \half \big( 1 \pm \gamma_\mu \big).}
If we now examine, for example, $\Gamma_2^\pm$ we see that
\eqn{\Gamma_2^\pm \begin{pmatrix} \psi^1 \\ \psi^2 \\ \psi^3 \\ \psi^4 \end{pmatrix} = 
\begin{pmatrix} \psi^1 \mp \psi^4 \\ \psi^2 \pm \psi^3 \\ \pm \psi^2 + \psi^3 \\ \mp \psi^1 + \psi^4 \end{pmatrix}.
}
Similar expressions for $\mu=1,3,4$ allow us to deduce that we only need to evaluate the action of the links on the upper half (in spinor space) of $\Gamma_\mu^\pm \psi_{x\mp\mu}$, as the lower components are equal to the upper components multiplied by $\pm 1$ or $\pm i$. In doing so we can halve the number of floating point multiplications needed in the evaluation of $D_{\rm w}$, and also reduce intermediate memory usage. This trick can be applied in any of the standard representations for the Euclidean-space $\gamma$ matrices.

\section{Generalised Spin-Proj\-ect\-ion Trick}

We now consider the case where there are two sets of links, $U_\mu(x)$ for the naive Dirac operator $\nabla\four$ and $U'_\mu(x)$ for the irrelevant Wilson term (denoted by $\Delta'$ to indicate that it contains only the links $U'$). In the case of a FLIC action the irrelevant links are APE-smeared, but what follows is perfectly general and does not depend upon any particular relationship between $U$ and $U'$. Now our ``split-link'' operator is
\begin{multline}
(D_{\rm split}\psi)_x = \big((\nabla\four + \frac{1}{2}\Delta' + m)\psi\big)_x\\
 = (4+m)\psi_x-\half\sum_\mu \big(U'_\mu(x)-\gamma_\mu U_\mu(x)\big)\psi_{x+\mu} \\
 + \big(U'{}^\dagger_\mu(x) +\gamma_\mu U^\dagger_\mu(x)\big)\psi_{x-\mu}.
\end{multline}
We can observe that our projectors do not present themselves immediately as they did before. At this point, compared to the standard Wilson action, we must perform four times as many floating point multiplications, two for the split links, and two for the loss of the spin projectors. However, we have
\eqn{{\bf 1} = \Gamma_\mu^+ + \Gamma_\mu^- \text{ and } \gamma_\mu =  \Gamma_\mu^+ - \Gamma_\mu^-, }
which implies
\begin{widetext}
\begin{multline}
(D_{\rm split}\psi)_x = (4+m)\psi_x-\half\sum_\mu \big(U'_\mu(x)-U_\mu(x)\big)\Gamma_\mu^+\psi_{x+\mu}
 + \big(U'_\mu(x)+U_\mu(x)\big)\Gamma_\mu^-\psi_{x+\mu} + \\ \big(U'{}^\dagger_\mu(x)+U^\dagger_\mu(x)\big)\Gamma_\mu^+\psi_{x-\mu}
 + \big(U'{}^\dagger_\mu(x)-U^\dagger_\mu(x)\big)\Gamma_\mu^-\psi_{x-\mu}.
\end{multline}
It is now clear that by defining symmetrised and anti-symmetrised links,
\eqn{ U^+_\mu(x) = \half\big(U'_\mu(x)+U_\mu(x)\big) \text{ and } U^-_\mu(x) = \half\big(U'_\mu(x)-U_\mu(x)\big) }
we can write
\eqn{
(D_{\rm split}\psi)_x = (4+m)\psi_x-\sum_\mu U^-_\mu(x)\Gamma_\mu^+\psi_{x+\mu} + U^+_\mu(x)\Gamma_\mu^-\psi_{x+\mu} +
U^{+\dagger}_\mu(x)\Gamma_\mu^+\psi_{x-\mu}+  U^{-\dagger}_\mu(x)\Gamma_\mu^-\psi_{x-\mu}. \label{eqn:splitspin}
}
\end{widetext}
Immediately we see that the Wilson spin projection trick is simply a special case of the split link trick where $U=U'$. The same saving in multiplications that we received in the Wilson case applies here, so we have in principle a factor of two compared to the Wilson action because $U^-$ is not zero. In actuality, efficient cache usage will reduce this to less than a factor of two.

\begin{widetext}
\section{The FLIC fermion action}

The FLIC action is a split-link action with clover term\cite{sheik-clover},
\eqn{
D_{\rm flic} = \nabla\four + \frac{1}{2}(\Delta' - \frac{c_{\rm sw}}{2}\sigma\cdot F) + m,
}
where 
\begin{eqnarray}
\sigma_{\mu\nu}&=&\frac{1}{2}[\gamma_\mu,\gamma_\nu], \quad F_{\mu\nu}(x) =\frac{1}{2}\big(C_{\mu\nu}(x)-C_{\mu\nu}^\dagger(x)\big), \\
C_{\mu\nu}(x)&=& \frac{1}{4}\big( U_{\mu\nu}(x) + U_{-\nu\mu}(x) + U_{\nu-\mu}(x) + U_{-\mu-\nu}(x) \big).
\end{eqnarray}
APE-smearing\cite{ape-one,ape-two, derek-smooth, ape-MIT} is carried out on the individual links in the irrelevant operators by making the replacement
\eqn{
\label{eqn:apesweep} U_\mu(x) \rightarrow U^{(\alpha)}_\mu(x) = {\mathcal P}\left( (\alpha -1) U_\mu(x) + \frac{\alpha}{6}\sum_{\pm\nu\neq\mu} U_\nu(x)U_\mu(x+ae_\nu)U^\dagger_\nu(x+ae_\mu) \right).
}
\end{widetext}
Here ${\mathcal P}$ denotes projection of the RHS of Eq.\ (\ref{eqn:apesweep}) back to the SU(3) gauge group. That is, each link is modified by replacing it with a combination of itself and the surrounding staples to give a set of ``fat links''.  The means by which one projects back to SU(3) is not unique. We choose an SU(3) matrix $U^{(\alpha)}_\mu(x)$ such that the gauge invariant measure ${\rm Re}{\rm Tr}(U^{(\alpha)}_\mu(x)X^\dagger_\mu(x))$ is maximal, where $X_\mu(x)$ is the smeared link before projection, that is $U^{(\alpha)}_\mu(x)\equiv {\mathcal P}X_\mu(x)$. As the process of APE-smearing removes short-distance physics, it is preferable to only smear the irrelevant operators.

Here $\alpha$ is the smearing fraction and $n_{\rm ape}$ is the number of smearing sweeps (\ref{eqn:apesweep}) we perform. Finally, as in \cite{zanotti-hadron}, we can perform tadpole or mean-field improvement (MFI) \cite{lepage-mfi} to bring our links closer to unity. This consists of updating each link with a division by the mean link, which is the fourth root of the average plaquette,

\eqn{ u_0 = \langle {\rm \frac{1}{3}ReTr } U_{\mu\nu}(x) \rangle_{x,\mu<\nu}^{\frac{1}{4}}. }

For completeness, we review a (well-known) similar trick for the clover term that exploits the structure of $\sigma_{\mu\nu}$. In the evaluation of the clover term, we note that in the chiral representation of $\gamma$ matrices, 
\eqn{ \gamma_4 = \begin{pmatrix} \bf 0 & \bf 1 \\ \bf 1 & \bf 0 \end{pmatrix} \quad   \gamma_5 = \begin{pmatrix} \bf 1 & \bf 0 \\ \bf 0 & \bf -1 \end{pmatrix}, }
the matrix $\sigma_{\mu\nu}$ satisfies the following  (in $2\times2$ block notation),
\begin{multline}
\sigma_{12}\begin{pmatrix} \psi \\ \chi \end{pmatrix} = \sigma_{34}\begin{pmatrix} -\psi \\ \chi \end{pmatrix},
\sigma_{13}\begin{pmatrix} \psi \\ \chi \end{pmatrix} = \sigma_{24}\begin{pmatrix} \psi \\ -\chi \end{pmatrix} \\ \text{ and }
\sigma_{14}\begin{pmatrix} \psi \\ \chi \end{pmatrix} = \sigma_{23}\begin{pmatrix} -\psi \\ \chi \end{pmatrix}.
\end{multline}
So we have, for example,
\eqn{
F_{12}\sigma_{12}\begin{pmatrix} \psi \\ \chi \end{pmatrix} + F_{34}\sigma_{34}\begin{pmatrix} \psi \\ \chi \end{pmatrix} = \begin{pmatrix} (F_{12} - F_{34})\sigma_{12}\psi \\ (F_{12}+ F_{34})\sigma_{12}\chi \end{pmatrix}.
}
This means that if we store the combinations $F_{12} \pm F_{34}, F_{13} \pm F_{24}, F_{14} \pm F_{23}$ we can halve the number of floating point multiplications needed in the evaluation of the clover term, further improving the computational efficiency of the FLIC action.

\section{conclusion}

We have presented a generalised version of the spin-projection trick which is applicable to any split-link action. This allows us to halve the number of floating-point multiplications the the evaluation of the action of the links upon the fermion field. We have also recalled some symmetries of $\sigma_{\mu\nu}$ in the chiral $\gamma$ matrix representation which allow us to perform a similar cost reduction in the evaluation of the clover term. The results presented here reduce the cost of evaluating the FLIC action to about twice that of the standard Wilson action. The exact difference will vary depending upon the base architecture, but on our architecture we have verified that the cost of FLIC is almost exactly twice that of the Wilson, including the cost of the clover term. Additionally, the formulation of the split link action in (\ref{eqn:splitspin}) allows groups who have efficient code for the Wilson action to simply implement efficient code for the FLIC action. Given the benefits of the FLIC action \cite{zanotti-hadron, kamleh-overlap} we hope that this work encourages groups to consider using the FLIC action for their calculations.

\begin{acknowledgments}
The author wishes to thank Urs Heller, Robert Edwards and Herbert Neuberger for discussions at Cairns in 2001 that led to this investigation. Additional thanks go to Patrick Bowman for helpful discussions immediately after Cairns, and to Tony Williams and Derek Leinweber for valuable discussions and their contribution to the manuscript. Thanks also go to David Richards for correspondence. This work was supported by the Australian Research Council.
\end{acknowledgments}



\begin{thebibliography}{10}

\bibitem{zanotti-hadron}
CSSM Lattice, J.~M. Zanotti {\em et~al.},
\newblock Phys. Rev. {\bf D65}, 074507 (2002), hep-lat/0110216.

\bibitem{kamleh-overlap}
W.~Kamleh, D.~H. Adams, D.~B. Leinweber, and A.~G. Williams,
\newblock Phys. Rev. {\bf D66}, 014501 (2002), hep-lat/0112041.

\bibitem{neuberger-practical}
H.~Neuberger,
\newblock Phys. Rev. Lett. {\bf { 81}}, 4060 (1998), hep-lat/9806025.

\bibitem{edwards-practical}
R.~G. Edwards, U.~M. Heller, and R.~Narayanan,
\newblock Nucl. Phys. {\bf { B540}}, 457 (1999), hep-lat/9807017.

\bibitem{alford-improving}
M.~G. Alford, T.~R. Klassen, and G.~P. Lepage,
\newblock Nucl. Phys. {\bf B496}, 377 (1997), hep-lat/9611010.

\bibitem{sheik-clover}
B.~Sheikholeslami and R.~Wohlert,
\newblock Nucl. Phys. {\bf B259}, 572 (1985).

\bibitem{ape-one}
M.~Falcioni, M.~L. Paciello, G.~Parisi, and B.~Taglienti,
\newblock Nucl. Phys. {\bf B251}, 624 (1985).

\bibitem{ape-two}
APE, M.~Albanese {\em et~al.},
\newblock Phys. Lett. {\bf B192}, 163 (1987).

\bibitem{derek-smooth}
F.~D.~R. Bonnet, D.~B. Leinweber, A.~G. Williams, and J.~M. Zanotti,
\newblock (2001), hep-lat/0106023.

\bibitem{ape-MIT}
M.~C. Chu, J.~M. Grandy, S.~Huang, and J.~W. Negele,
\newblock Phys. Rev. {\bf D49}, 6039 (1994), hep-lat/9312071.

\bibitem{lepage-mfi}
G.~P. Lepage and P.~B. Mackenzie,
\newblock Phys. Rev. {\bf D48}, 2250 (1993), hep-lat/9209022.

\end{thebibliography}

\end{document}